\documentclass[aps,preprint,showpacs,preprintnumbers,floatfix]{revtex4}
\usepackage{graphicx}% Include figure files 
\usepackage{dcolumn}%Align table columns on decimal point 
\usepackage{bm}% bold math

\begin{document}

\preprint{LBNL-51306}

\title{Liquid-vapor phase transition in nuclei \\
or compound nucleus
decay? }

\author{L.G. Moretto, J.B. Elliott, L. Phair, G.J. Wozniak}
\affiliation{Nuclear Science Division, Lawrence Berkeley National
Laboratory \\ University of California, Berkeley, California 94720 }
\homepage{http://www.lbl.gov/~phair} \email{lgmoretto@lbl.gov}

\date{\today}

\begin{abstract}
Recent analyses of multifragmentation in terms of Fisher's model and
 the related construction of a phase diagram brings forth the problem
 of the true existence of the vapor phase and the meaning of its
 associated pressure.  Our analysis shows that a thermal emission picture is 
 equivalent to a Fisher-like equilibrium description which avoids the
 problem of the vapor and explains the recently observed
 Boltzmann-like distribution of the emission times. In this picture a simple Fermi gas
 thermometric relation is naturally justified. Low energy compound
 nucleus emission of intermediate mass fragments is shown to scale
 according to Fisher's formula and can be simultaneously fit with the
 much higher energy ISiS multifragmentation data.
\end{abstract}


\maketitle

%\section{\label{sec:level1}First-level heading:\protect\\ The line
%break was forced \lowercase{via} \textbackslash\textbackslash}

After many decades of theoretical studies and of experimental
pre-discoveries, recent papers have published what can be considered a
quantitative, credible liquid-vapor phase diagram containing the
coexistence line up to the critical temperature \cite{Elliott:ISiS}. Somewhat
unexpectedly, this diagram has not been obtained through the study of
caloric curves \cite{Pochodzalla:caloric_curve,Chernomoretz:caloric_curve} or anomalous heat capacities
\cite{Gulminelli:critical_behavior,Dagostino:neg_C}. Rather, it was generated from the fitting of the
charge distributions in multifragmentation by means of a Coulomb
corrected Fisher's formula \cite{Elliott:ISiS,Fisher:droplet69} giving the cluster composition of a vapor:
\begin{equation}
n_A(T)=q_0A^{-\tau}\exp\left[\frac{\Delta\mu A}{T}-\frac{c_0\epsilon
A^{\sigma}}{T}-\frac{E_{\rm Coul}}{T}\right],
\label{eq:Fisher}
\end{equation}
where $q_0$ is a normalization constant \cite{Fisher:droplet69}, $\tau$ is the critical exponent giving rise to a power law at
criticality, $A$ is the cluster number, $\Delta\mu$ is the difference
of chemical potentials between the liquid and the vapor, $c_0$ is the
surface energy coefficient, $T$ is the temperature, $\epsilon$ is the distance from the critical temperature $T_c$ and is $\epsilon=(T_c-T)/T_c$, $\sigma$ is
another critical exponent (expected to be approximately 2/3, if one interprets the
second term in the exponent as the surface energy of a cluster of mass
$A$ divided by the temperature) and $E_{\rm Coul}$ is the Coulomb energy \footnote{A specific form for the Coulomb term was used in \cite{Elliott:ISiS} which accounts for the change in sign in the notation of this paper and that of \cite{Elliott:ISiS}.}.
For $\Delta\mu=0$ the liquid and the vapor are in equilibrium and
Eq.~(\ref{eq:Fisher}) can be taken to be the equivalent of the
coexistence line. More conventionally, one can immediately obtain from
Eq.~(\ref{eq:Fisher}) the usual $p,T$ and $\rho ,T$ phase diagrams
by recalling that in Fisher's model, the clusterization is assumed to exhaust all the
non-idealities of the gas. It then becomes an ideal gas of
clusters. Consequently, the total pressure is
\begin{equation}
p(T)=\sum _A p_A(T)=T\sum _A n_A(T),
\label{eq:tot_p}
\end{equation}
the scaled pressure $p/p_c$ is
\begin{equation}
\frac{p}{p_c} = \frac{T\sum n_A(T)}{T_c\sum n_A(T_c)},
\label{eq:scaled_p}
\end{equation}
and the density is
\begin{equation}
\rho=\sum _A An_A(T).
\end{equation} 

Tests on the 3-dimensional Ising model \cite{Mader:Ising} demonstrate a
beautiful agreement between the Ising cluster distributions and
Eq.~(\ref{eq:Fisher}), and analysis of many multifragmentation
reactions \cite{Elliott:ISiS,Elliott:EOS2002} show equally good agreement, leading to
the claim of characterization of the nuclear liquid-vapor phase diagram.

The only troubling point in this otherwise elegant picture is
summarized by the question: where is the vapor?
Does the nuclear system truly present itself at
some time like a mixed phase system with the vapor being somehow
restrained, either statically or dynamically in contact with the
liquid phase, whatever that might be? And, what is the meaning of vapor
pressure, when clearly the system is freely decaying in vacuum against
no pressure?

The purpose of this paper is to show: 

$\bullet$ why an
equilibrium description, such as Fisher's, is relevant to the free
vacuum decay of a multifragmenting system; 

$\bullet$ how we can talk about
coexistence without the vapor being present;

$\bullet$ and why a simple
thermometric equation such as $E=aT^2$ works better than empirical
thermometers such as isotope thermometers. 

%$\bullet$ Why caloric curves and
%(negative) heat capacities may be misleading tools with regards to
%phase transitions?

%There are additional points which will be touched upon as we progress.

We begin with a time-honored assumption which we do not try to justify
other than through the clarification it brings to the experimental
picture.
We assume that, after prompt emission in the initial phase of the collision has been isolated or accounted for, the resulting system relaxes
in shape and density and thermalizes {\em on a time scale faster than its
thermal decay}. This will undoubtedly bring to mind the compound nucleus assumption, and not without reason.

At this point the system emits particles in vacuum, according to
standard statistical decay rate theories. Experimentally, the initial excitation
energy is typically evaluated calorimetrically after
accounting for pre-equilibrium emission, and the initial
temperature can be estimated by the thermometric equation of a Fermi
gas
\begin{equation}
E = aT^2
\end{equation}
allowing perhaps for a weak dependence of $a$ on $T$, and remembering that
the system is most likely still in the strongly degenerate regime.

But again, what is the relevance of this to liquid-vapor phase transition, and where
is the vapor?

Let us for a moment imagine the nucleus surrounded with its saturated vapor. At
equilibrium, any particle evaporated by the nucleus will be restored
by the vapor bombarding the nucleus. In other words, the outward
evaporation flux from the nucleus to the vapor is exactly matched by
the inward condensation flux. This is true for any kind of evaporated
particle. Thus, the vapor acts like a mirror, reflecting back into the
nucleus the particles which it is trying to evaporate. One can
obviously probe the vapor by putting a detector in contact with
it. But since the outward and inward fluxes are identically the same,
one might as well put the detector in contact with the nucleus
itself. At equilibrium, the two measured fluxes must be the
same. Therefore, we do not need the vapor to be present in order to
characterize it completely.  We can just as well study the evaporation
of the nucleus in equilibrium and dispense with our imaginary surrounding saturated vapor.

Quantitatively, we can simply relate the concentration  $C_A(T)$ of any species
$A$ in the vapor to the corresponding decay rate $P_A(T)$ (controlled by a decay width $\Gamma _A$) from the
nucleus by matching the fluxes
\begin{equation}
P_A(T) = \frac{\Gamma_A(T)}{\hbar} =C_A(T)\left<v_A(T)\sigma _{\rm inv}(v_A)\right>
\label{eq:rate}
\end{equation}
where $v_A(T)$ is the velocity of the species $A$ (of order
$(T/A)^{1/2}$) crossing the nuclear interface represented by the inverse cross section $\sigma _{\rm inv}$.

Thus, the vapor phase in equilibrium can be completely characterized
in terms of the decay rate. The vapor need not be there at all. This
is not a nuclear peculiarity. It is just the same for a glass of water
exposed to dry air or vacuum. One speaks in these situations of a
``virtual vapor'', realizing that first order phase transitions depend
exclusively upon the intrinsic properties of the two phases, and not
on their interaction. But, of course, if the vapor is not there to restore the emitting system with its back flux,
evaporation will proceed, leading to a cooling off of the
system. Instantaneously, the physical picture described above is still valid, but not
globally. The result of a global evaporation in vacuum is unfortunate
in terms of the analysis, as it integrates over a continuum of
temperatures. It is unfortunate for the complications it lends to the
possible thermometers (kinetic energy, isotope ratios, etc.), as well
as to the abundances of the various species. In this aspect lies the real difference between our approach and any true equilibrium approach.

But, there is a simple, astute way to avoid this complication. Let us choose
to consider only particles that are emitted very rarely so that, if
they are not emitted at the beginning of the decay, they are effectively not
emitted at all. In other words, let us consider only particles that by
virtue of their large surface energy, have a high emission
barrier. 

As an example, consider a decaying system with only three available exit channels. We call them channels $a, b, $ and $n$ with barriers $B_a$, $B_b$, and $B_n$. For $B_n \ll B_a$ and $B_b$ we know that the probability of emission of particles of type $b$ at a fixed temperature is approximately
\begin{equation}
p_b \approx e^{-(B_b-B_n)/T}.
\end{equation}
Since the nucleus cools as particles are emitted, the total emission probability of particles of type $b$ from a nucleus at initial temperature $T_0$ goes like
\begin{equation}
P_b\propto\int _0 ^{T_0} e^{-(B_b-B_n)/T}2aT dT.
\end{equation} 
A similar expression exists for $P_a$. The ratio of $P_b/P_a$ is
\begin{equation}
\frac{P_b}{P_a}=\frac{\Delta _b^2}{\Delta _a^2}
\frac{\int _0^{T_0/\Delta _b}e^{-1/x}x dx}{\int _0^{T_0/\Delta _a}e^{-1/x}x dx}
\end{equation}
where $\Delta _b=B_b-B_n$ and $\Delta _a=B_a-B_n$. 
The ratio $P_b/P_a$ can also be used to extract an effective temperature $T_{\rm eff}$
\begin{equation}
\frac{P_b}{P_a}=\exp\left(-\frac{B_b-B_a}{T_{\rm eff}}\right).
\end{equation}

\begin{figure}
% from /home/lwphair/Tave
\includegraphics[angle=90,width=8.0cm]{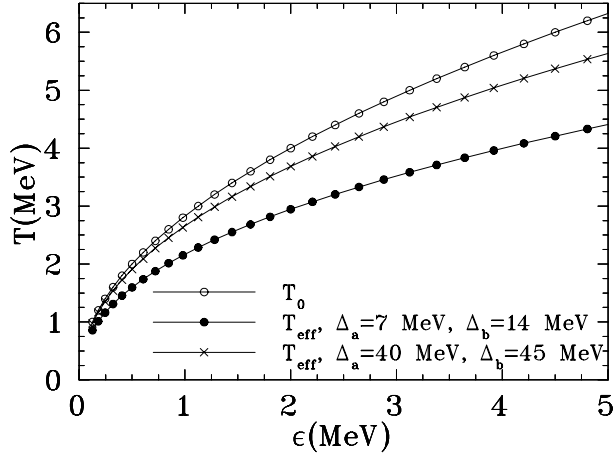}
\caption{\label{fig:cooling} The effective temperature of a Fermi system with three exit channels ($a$, $b$, and $n$) is plotted as function of initial excitation energy for two cases: one where barriers $B_a$ and $B_b$ are large (crosses) compared to $B_n$=6 MeV, and another where $B_a$ or $B_b$ is similar (solid circles) to $B_n$. The initial temperature as a function of initial excitation energy is shown by the open circles.}
\end{figure}

An example of how the effective temperature compares with the initial temperature $T_0$ is given in Fig.~\ref{fig:cooling} for different values of $B_b$ and $B_a$. The case where $B_a$ and $B_b$ are large (crosses) gives effective temperatures very near to the initial temperature $T_0$ (open circles). When either $B_a$ or $B_b$ is near the barrier of the most probable channel (solid circles), the effective temperature is very different from the initial temperature. 

Our goal then should be to choose exit channels with large barriers in order to justify our use of the initial Fermi temperatures. 
This is what has been done 
%(perhaps a bit serendipitously) 
in
the analyses leading to the nuclear phase diagrams \cite{Elliott:ISiS,Elliott:EOS2002}, where the
fragments with charge $Z < 5$ were excluded. Under these conditions, the
validity of Eq.~(\ref{eq:rate}) is guaranteed. The rate can be related to the
vapor concentration and the phase diagram can be constructed. The
temperature necessary for our purpose is fortunately the initial
temperature and not the average temperature determined for multiply emitted
particles. The correctness of a thermometric relation $E = aT^2$ can be
tested ``a posteriori'' by verifying the linearity of the Fisher's
plots \cite{Elliott:ISiS,Elliott:EOS2002} and their predecessors 
\cite{Moretto:reducibility_review}. This linearity, extending over many orders of magnitude for a
variety of fragments, is in our view the strongest test yet of a
Fermi gas thermometric relationship. In fact one can turn the problem around and determine the thermometric relationship up to rather high excitation energies by the requirement that it leads to a linear Fisher's plot.

We offer three additional proofs for our physical picture of a hot remnant evaporating particles.  

\begin{figure}
\includegraphics[angle=90,width=8.0cm]{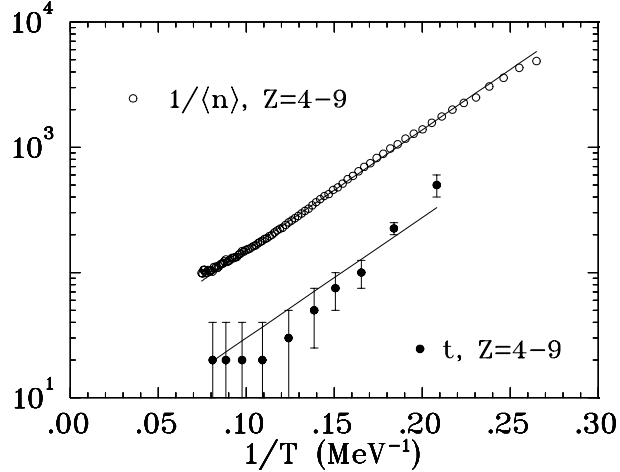}
\caption{\label{fig:mean_time} The mean emission times (in fm/c) of fragments with atomic
number $4\le Z\le 9$ are plotted (solid symbols) versus inverse temperature for the reaction
$\pi$+Au at 8 GeV/c \protect\cite{Beaulieu:imftimes,Beaulieu:reducibility_ISiS}. The average yields  of the same fragments are plotted versus $1/T$ (solid symbols). The line
represents a Boltzmann fit to the fragment yields. This same line has been superimposed (shifted) on to the emission times.}
\end{figure}

First, the abundances of
the observed fragments as a function temperature allow us to construct an
Arrhenius plot ($\log P$ versus $1/T$) which is equivalent to a Fisher's
plot \cite{Elliott:ISiS,Elliott:EOS2002,Moretto:reducibility_review}. The slope is the effective ``barrier'' $B$ for the emission of the
particle.
This can be seen immediately by considering that the yields $\left< n\right>$ reflect the thermal scaling of the decay width
\begin{equation}
\left<n\right>\propto\Gamma\propto e^{-B/T}.
\end{equation}
But the very same barrier and the very same Boltzmann factor
intervene in determining the mean time separation $t$ between two
fragments since
\begin{equation}
t=\frac{\hbar}{\Gamma}\propto e^{B/T}.
\end{equation}
Such a time $t$ is the reciprocal of $\Gamma$. Therefore, the
same Arrhenius plot with the same barrier ought to explain both
the temperature dependence of the abundances and of the times. This is
exactly the case as shown in Fig.~\ref{fig:mean_time}. The ISiS collaboration has measured the yields (open symbols) \cite{Elliott:ISiS} and the mean emission times (solid symbols) \cite{Beaulieu:imftimes,Beaulieu:reducibility_ISiS} of intermediate mass fragments as a function of excitation energy. These energies can be translated into a Fermi temperature \cite{Elliott:ISiS} as discussed above. A Boltzmann fit to the yields is shown by the solid line. That same line has been superimposed (shifted) onto the emission time data and describes the data very well. In other words, the two different observables and their energy dependence are described by the same barrier.

Second, since all that has been said above holds exactly for low excitation energies, compound nuclear decay suddenly becomes relevant to the liquid-vapor phase transition. We should be able to scale known low energy
compound nucleus particle yields \cite{Fan:Se} according to the Fisher's scaling. 

\begin{figure}
\includegraphics[width=6.0cm,angle=90]{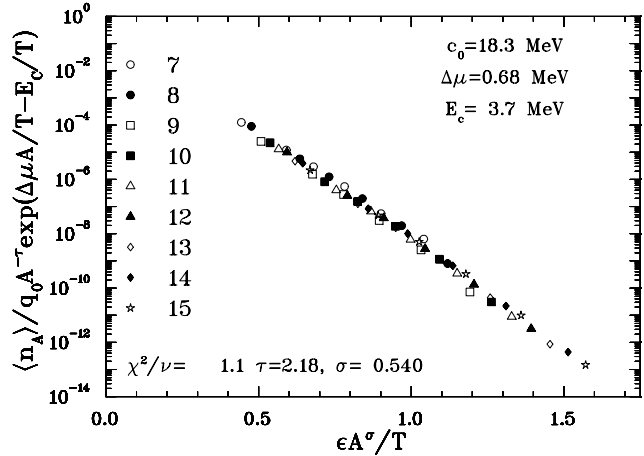}
\caption{\label{fig:cn} Results for the Fisher-scaled yield distribution
	versus the scaled temperature for the Ni$+$C low energy
	compound nucleus decay data.}
\end{figure}

This works out rather well as can be seen in Fig.~\ref{fig:cn} for the
reaction of $^{64}$Ni+$^{12}$C \cite{Fan:Se}. These data were taken at
the 88-inch cyclotron using Ni beams with energies between 6 and 13
MeV/nucleon. Given that the excitation energies are extremely small and that the fragment emission barriers are large compared to those of neutron evaporation,
there is here little doubt about a thermometric relation of the kind
$E=aT^2$. The data have been scaled using the very same Fisher
parameters as extracted from the ISiS data \cite{Elliott:ISiS}, except
for the critical excitation energy $E_C$, the Coulomb correction parameter \cite{Elliott:ISiS,Elliott:EOS2002}, and the value of $\Delta\mu$ which were allowed to vary freely. The
values of the Fisher parameters are listed in Fig.~\ref{fig:cn}. 
%The
%surface energy coefficient is only 0.5 MeV different from that
%determined by the high energy ISiS data \cite{Elliott:ISiS}. 

The data
scale over many orders of magnitude. With the compound nucleus data,
we are far from the critical temperature, yet the resulting extraction
of $E_C$ gives only a modest uncertainty ($\pm 0.3$ MeV). If the other
Fisher parameters are also allowed to vary freely (not constrained to
ISiS values), the uncertainty of $E_C$ becomes large, $\pm 2$
MeV. Still, it is remarkable that we observe a consistent
scaling in the compound nucleus data using the scaling parameters from
% but we also recover nearly the
%same surface energy coefficient and critical excitation energy as in
the high excitation energy experiments. From this example we see in
these low energy reactions a very interesting source for further
characterization of the phase transition, in particular for anchoring
the parameters of Fisher's model to the well established $T$=0
parameters of the liquid drop model.

\begin{figure}
\includegraphics[width=9.0cm]{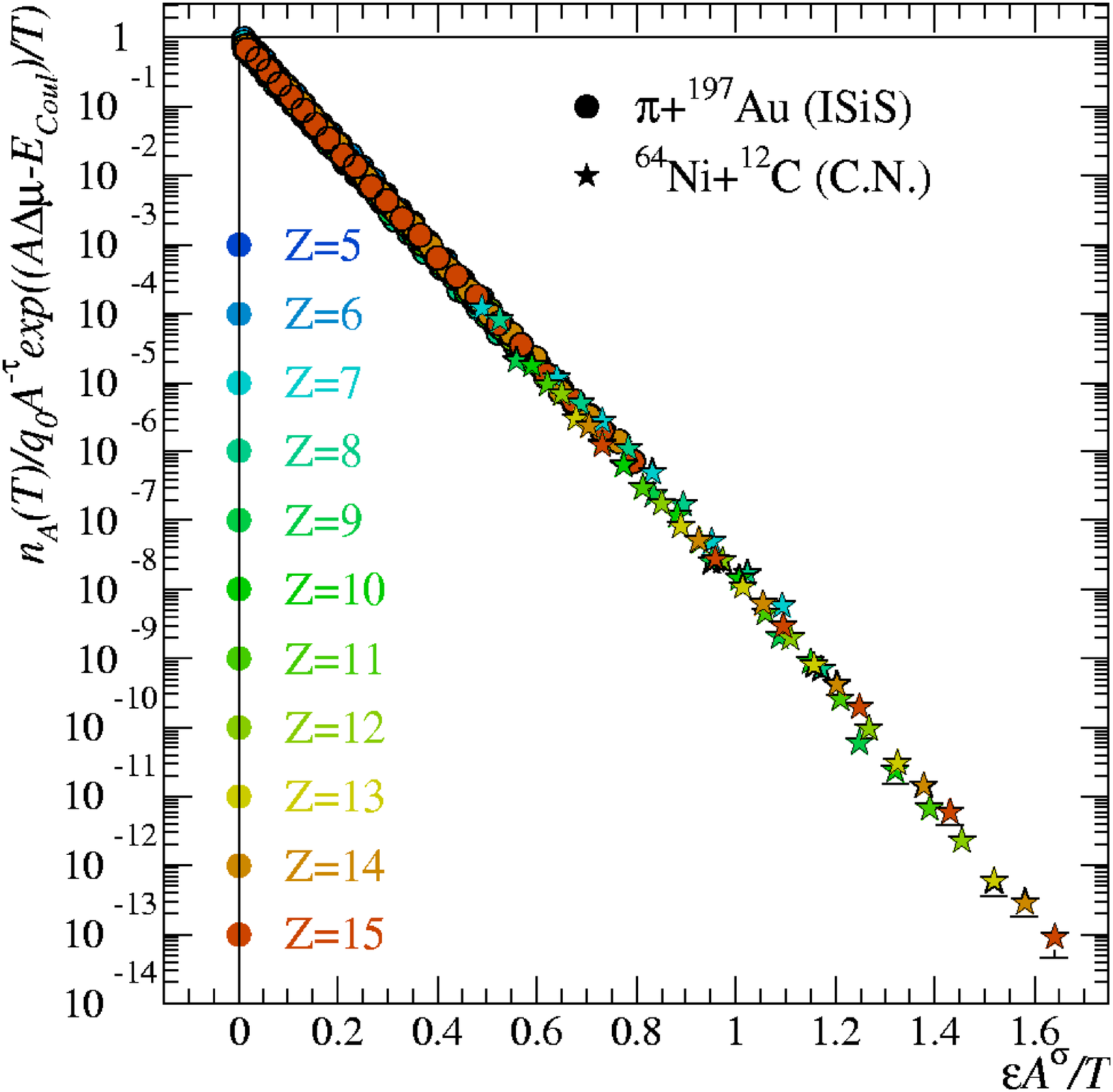}
\caption{\label{fig:all} The Fisher scaled yields are plotted versus the scaled temperature for the indicated reactions.}
\end{figure}
  
For our third and final demonstration, we show the results of a consistent fit of the ISiS data \cite{Elliott:ISiS}
% the EOS data \cite{Elliott:EOS2002}, 
and of the low energy compound nucleus data \cite{Fan:Se} with the Fisher model modified for Coulomb (Eq.~(\ref{eq:Fisher})). The resulting Fisher scaling is shown in Fig.~\ref{fig:all} for both systems. A smooth, continuous behavior is observed from the compound nucleus data up to the higher energy systems. This smooth behavior using a consistent set of Fisher parameters indicates a natural extension of the compound nuclear decay mechanism up to higher energies.

In conclusion, the ISiS 
%and EOS
data as well as low energy compound nucleus
data contain the signature of a liquid to vapor phase transition via
their strict adherence to Fisher's model. Through a direct examination
of the mean emission times of the ISiS fragmentation reactions, we
infer a stochastic, thermal emission scenario consistent with complex fragment
emission at much lower excitation energies. 

\begin{acknowledgments}
This work was supported by the US Department of Energy.
\end{acknowledgments}

%\section{TO DO}
%New figures with better labels.
% New title?
% List e-print address for Cathy's paper

%\thebibliography{99}%
\bibliography{mybib} 
%\bibitem{Elliott:ISiS} J.B. Elliott {\em et al.}, Phys. Rev. Lett. {\bf 88}, 047201 (2002).
%\bibitem{Pochodzalla:caloric_curve} J. Pochodzalla {\em et al.}, Phys. Rev. Lett. {\bf 75}, 1040 (1995).
%\bibitem{Chernomoretz:caloric_curve} A. Chernomoretz {\em et al.}, Phys. Rev. C {\bf 64}, 044605 (2001).
%\bibitem{Gulminelli:critical_behavior} F. Gulminelli and Ph. Chomaz, Phys. Rev. Lett. {\bf 82}, 1402 (1999).
%\bibitem{Dagostino:neg_C} M. D'Agostino {\em et al.}, Phys. Lett. B {\bf 473}, 219 (2000).
%\bibitem{mader} C.M. Mader {\em et al.}, preprint LBNL-47575.
%
%\bibitem{Elliott:EOS2002} J.B. Elliott {\em et al.}, submitted to Phys. Rev. C, LBNL-XX.
%\bibitem{Moretto:reducibility_review} L.G. Moretto {\em et al.}, Phys. Rep. {\bf 287}, 249 (1997).
%\bibitem{Beaulieu:imftimes} L. Beaulieu {\em et al.}, Phys. Rev. Lett. {\bf 84}, 5971 (2001).
%\bibitem{Beaulieu:reducibility_ISiS} L. Beaulieu {\em et al.}, Phys. Rev. C {\bf 63}, 031302 (2001).
%
%\bibitem{Fan:Se} T.S. Fan {et al.}, Nuc. Phys. A {\bf 679}, 121 (2000).

\end{document}